\documentclass[12pt]{article}
\usepackage{pic04}
\usepackage{amsmath}
\usepackage{hyperref}
\usepackage{url}
\usepackage{graphicx}

\def\Journal#1#2#3#4{{#1} {\bf #2} (#3) #4}

\def\NPB{{Nucl. Phys.} B}
\def\PLB{{Phys. Lett.} B}
\def\PRL{Phys. Rev. Lett.}
\def\PRD{{Phys. Rev.} D}
\def\ZPC{{Z. Phys.} C}
\def\EPJ{{Eur. Phys. J.} C}

\newcommand{\kspll}{K_{S}\rightarrow{\pi^0}l^+l^-}
\newcommand{\klpll}{K_{L}\rightarrow{\pi^0}l^+l^-}
\newcommand{\ksspll}{K_{1}\rightarrow{\pi^0}l^+l^-}
\newcommand{\kspee}{K_{S}\rightarrow{\pi^0}e^+e^-}
\newcommand{\kspmm}{K_{S}\rightarrow{\pi^0}\mu^+\mu^-}
\newcommand{\klpee}{K_{L}\rightarrow{\pi^0}e^+e^-}
\newcommand{\kcpen}{K^{+}\rightarrow{\pi^0}e^+\nu}
\newcommand{\kleegg}{K_{L}\rightarrow e^+e^-\gamma\gamma}
\newcommand{\klpmm}{K_{L}\rightarrow{\pi^0}\mu^+\mu^-}
\newcommand{\klpgg}{K_{L}\rightarrow{\pi^0}\gamma\gamma}
\newcommand{\kpnn}{K\rightarrow{\pi}\nu\overline{\nu}}
\newcommand{\klpnn}{K_L\rightarrow{\pi^0}\nu\overline{\nu}}
\newcommand{\kcpnn}{K^{+}\rightarrow{\pi^+}\nu\overline{\nu}}
\newcommand{\klpp}{K_{L}\rightarrow{\pi^+}\pi^-}
\newcommand{\klppn}{K_{L}\rightarrow{\pi^0}\pi^0}
\newcommand{\klppp}{K_{L}\rightarrow{\pi^0}\pi^0\pi^0}
\newcommand{\kspp}{K_{S}\rightarrow{\pi^+}\pi^-}
\newcommand{\ksppn}{K_{S}\rightarrow{\pi^0}\pi^0}
\newcommand{\klmm}{K_{L}\rightarrow{\mu^+}\mu^-}
\newcommand{\klsmm}{K_{L,S}\rightarrow{\mu^+}\mu^-}
\newcommand{\ksppp}{K_{S}\rightarrow{\pi^0}\pi^0\pi^0}
\newcommand{\kspgg}{K_{S}\rightarrow{\pi^0}\gamma\gamma}
\newcommand{\ksgg}{K_{S}\rightarrow\gamma\gamma}
\newcommand{\klgg}{K_{L}\rightarrow\gamma\gamma}
\newcommand{\klgsg}{K_{L}\rightarrow\gamma^*\gamma}
\newcommand{\klgsgs}{K_{L}\rightarrow\gamma^*\gamma^*}
\newcommand{\kleeg}{K_{L}\rightarrow e^+e^-\gamma}
\newcommand{\kleeee}{K_{L}\rightarrow e^+e^-e^+e^-}
\newcommand{\kleemm}{K_{L}\rightarrow e^+e^-\mu^+\mu^-}
\newcommand{\klmmg}{K_{L}\rightarrow \mu^+\mu^-\gamma}
\newcommand{\gaga}{\gamma\gamma}
\newcommand{\kpln}{K \rightarrow \pi l \nu}
\newcommand{\kspen}{K_{S}\rightarrow \pi e \nu}

\newcommand{\klpen}{K_{L}\rightarrow \pi e \nu}
\newcommand{\klpeng}{K_{L}\rightarrow \pi e \nu \gamma}
\newcommand{\klppen}{K_{L}\rightarrow \pi^0 \pi e \nu}
\newcommand{\klpmn}{K_{L}\rightarrow \pi \mu \nu}
\newcommand{\klpppc}{K_{L}\rightarrow \pi^+\pi^-\pi^0}

\newcommand{\kpmpppc}{K^{\pm}\rightarrow \pi^{\pm}\pi^+\pi^-}
\newcommand{\kpmpppn}{K^{\pm}\rightarrow \pi^{\pm}\pi^0\pi^0}
\newcommand{\kpppp}{K^{+}\rightarrow \pi\pi\pi}
\newcommand{\kmppp}{K^{-}\rightarrow \pi\pi\pi}

\newcommand{\cpt}{{\chi}PT}

\begin{document}

\title{\bf KAON PHYSICS}
\author{
Ivan Mikulec        \\
{\em \"Osterreichische Akademie der Wissenschaften, Institut f\"ur
  Hochenergiephysik,} \\ 
{\em A-1050 Wien, Austria}}
\maketitle

%
% photograph of author
%  This is where we will insert a photograph. To see what it would look like,
%  uncomment the following lines.
%
%\begin{figure}[h]
%\begin{center}
%
% include photograph for proceeding version
%
%\includegraphics
%[height=4.5cm]{einstein.eps}
%
% insert a fixed vertical spacing instead for the ArXiv preprint
%
\vspace{4.5cm}
%
%\end{center}
%\end{figure}

\baselineskip=14.5pt
\begin{abstract}
The most recent progress and future prospects in kaon physics are
reviewed. Main results are
the first observation of the rare decays $\kspee$ and $\kspmm$ by
NA48, new $\kcpnn$ event observed by E949
and new precision measurements of $K_{S,L} \rightarrow \pi e \nu$
decays by KTeV, KLOE and NA48 made in an effort to 
resolve the slight deviation from the unitarity in
the CKM matrix, involving the $|V_{us}|$ coupling. 
\end{abstract}
\newpage

\baselineskip=17pt

\section{Introduction}

The observation and understanding of
kaon decays has been crucial for the progress in the theory of particle
physics, especially for the Standard Model (SM). The first observation of the
decay $\klpp$ by Christenson, Cronin, Fitch and Turlay 40 years 
ago~\cite{disc} has
shown for the first time that the CP symmetry is violated in
nature and guided Kobayashi and Maskawa, a decade later, to introduce
a third quark family into their quark mixing scheme~\cite{ckm}. 
Low decay rates observed 
in the decay $\klmm$, lead to the discovery of the GIM 
mechanism~\cite{gim} and to the
prediction of the charm quark. A few years ago, the direct
violation of the CP symmetry in the decay amplitude,
predicted by the Standard Model and first observed by NA31 with
3$\sigma$ evidence~\cite{na31}, 
has been firmly established by NA48~\cite{na48a,na48b} and 
KTeV~\cite{ktev}. At
present, the field of kaon physics still provides exciting results. 

In this paper, the subject is divided into three parts. In the first
part, the recent progress in the understanding of the rare decays
involving flavor-changing neutral currents (FCNC) is described. The
second part is devoted to semileptonic kaon decays especially to the
decay $K_{S,L} \rightarrow \pi e \nu$ or $K_{e3}$, the best probe for
the measurement of the CKM matrix element $|V_{us}|$. In the third
part, recent studies of CP violation in kaon decays are reviewed.

\section{Rare Decays}

In the quest to find New Physics (NP) beyond SM, the searches for
SM-forbidden kaon decays, in particular those which violate
the lepton flavor,
are ceding territory to measurements in the quark flavor sector.
GIM-suppressed one-loop FCNC processes are very sensitive to
additional flavor structure with up to
an order of magnitude enhancements with respect to SM decay rates 
predicted by various NP models~\cite{np,npa}. 

\begin{figure}[htb]
\includegraphics[width=13cm]{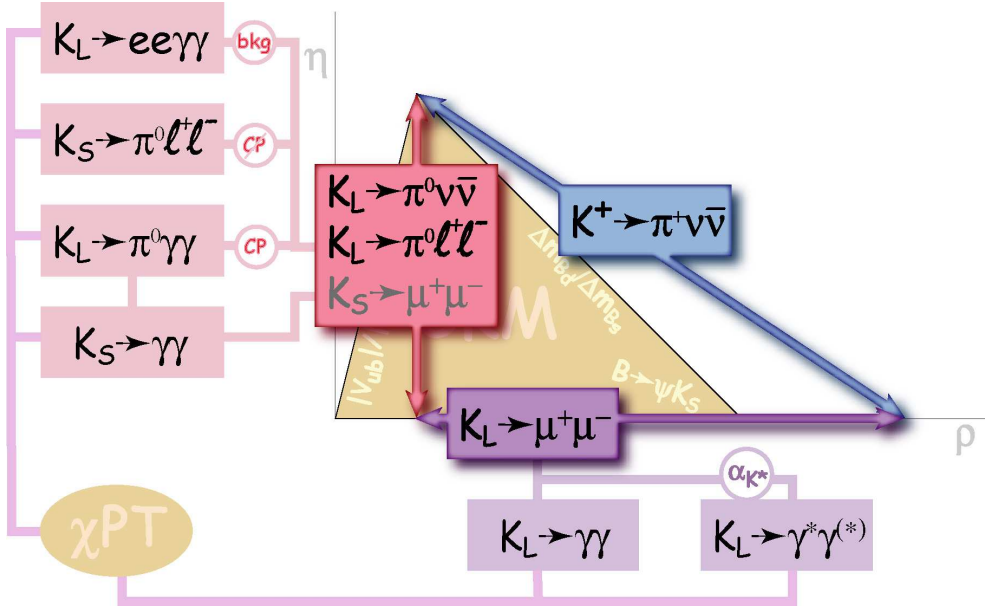}
 \caption{\it
    Rare kaon FCNC reactions can fully determine the CKM unitarity
    triangle. Long-distance contributions to some of these processes
    are calculated in the framework of $\cpt$ with the help of other,
    related kaon rare decays.
    \label{ckmtr} }
\end{figure}

There are two classes of these decays (Fig.~\ref{ckmtr}):
\begin{itemize}
\item Pure FCNC decays: $\klpnn$ and $\kcpnn$. 
  These decays are fully dominated by short-distance dynamics which
  can be calculated perturbatively and the hadronic matrix element can
  be obtained from the well measured $\kcpen$ decay.
  This leads to an extremely precise (to few \%) relation between
  decay rates and the combination of CKM matrix elements
  $\lambda_t=V^*_{ts}V_{td}$. This level of theoretical cleanness 
  in constraints to the CKM unitarity triangle is
  only matched by the CP asymmetry in $B \rightarrow J/\psi
  K_S$. Unfortunately, from the experimental point of view these decays are
  very challenging.
\item Decays with both short-distance and long-distance
  contribution: $\klpll$ ($l=e,\mu$) and $\klsmm$. Here, the long-distance
  amplitudes must be determined to a sufficient precision in order to access
  the short-distance physics. The determination of the long-distance
  contributions is done in the framework of Chiral Perturbation
  Theory ($\cpt$) which is an effective field theory of SM describing
  the low energy (at kaon mass) hadron dynamics. As $\cpt$ is an
  effective theory, many parameters have to be determined
  experimentally with the help of other rare kaon decays. 
\end{itemize}

\subsection{$\klpll$}

The decay amplitude of $\klpnn$, entirely due to direct CP
  violation~\cite{litt}, is proportional to $Im \lambda_t$ and determines the
  height of the CKM unitarity triangle usually denoted by the Wolfenstein
  parameter $\eta$~\cite{wolf}. 
The present experimental limit~\cite{klpnn} is many orders of
  magnitude above the SM expectation~\cite{isid}.

  While the experimental state of the art is slowly
  reaching the level of being capable to seriously search for this
  challenging decay, an alternative decay accessing the same physics
  is under study. This is the decay $\klpll$. The advantage of this
  decay is that all the decay products can be detected, though the
  presence of radiative background can become a significant
  obstacle. 

The short-distance direct CP-violating (DCPV) contribution
to the decay $\klpee$  calculated by
  Buchalla, D'Ambrosio and Isidori~\cite{bdi} is
\begin{equation}
BR(\klpee)_{DCPV} \approx 4.4 \times 10^{-12}
\end{equation}
and to the decay $\klpmm$ obtained by
  Isidori, Smith and Unterdorfer~\cite{isu} is
\begin{equation}
BR(\klpmm)_{DCPV} \approx 1.8 \times 10^{-12}
\end{equation}
in both cases using the latest SM CKM fit result 
$Im \lambda_t = 1.36\times 10^{-4}$~\cite{batt}. 

Unlike $\klpnn$, these decays
receive also two long-distance contributions. The CP-conserving (CPC)
contribution proceeds through two virtual photons and has been 
recently determined by a precise study of the decay $\klpgg$~\cite{klpgg}
in the low
invariant $\gaga$ mass region. The $BR(\klpee)_{CPC}$ is negligible~\cite{bdi}
while in the muon channel the CP-conserving contribution is
\begin{equation}
BR(\klpmm)_{CPC} \approx 5.2 \times 10^{-12}
\end{equation}
which is few times larger than the DCPV part~\cite{isu}.

The CP-conserving decay $\ksspll$, through $K^0-\overline{K^0}$ mixing,
provides an extra long-distance indirect CP-violating (ICPV)
contribution to the $\klpll$. In addition, DCPV and ICPV components
interfere with each other. Clearly, in order to determine this
long-distance component the knowledge of $BR(\kspll)$ is
necessary. The theoretical predictions, scattered over an order of
magnitude around few $10^{-9}$, do not provide a sufficient
unambiguity~\cite{theoks}. 

The search for $\kspee$ and $\kspmm$ decays was the primary goal of the
NA48/1 extension~\cite{add2} 
of the scientific program of the NA48 experiment at
CERN, taking place in the year 2002. 
This experiment, after successfully finishing the precise
measurement of $Re\ \varepsilon'/\varepsilon$ in 2001, took the opportunity to
use the well functioning apparatus to study rare $K_S$ decays. For
this purpose the original beam setup, consisting of simultaneous 
high-intensity far-target and low-intensity near-target beams from SPS, was
modified to run only the near-target beam line but at much higher
intensity. This setup has been successfully 
tested and employed during the year 2000, in a
period in which the spectrometer was in repair, leading to a series of
results on $K_S$ decays with purely photons in final state: 
$\ksgg$~\cite{ksgg}, $\kspgg$~\cite{kspgg},
$\ksppp$~\cite{ksppp}.

Both $\kspee$ and $\kspmm$ decays were successfully found by
NA48/1. In the $\kspee$ mode, 7 events with background expectation of 
$0.15^{+0.10}_{-0.04}$ events were found~\cite{kspee}, 
while in the $\kspmm$ mode, 6
events with $0.22^{+0.19}_{-0.12}$ of expected background were
observed~\cite{kspmm}. In the electron channel, in order to obtain a 
low background level a cut on $ee$ invariant mass, 
$m_{ee}>0.165$~GeV/$c^2$, had to be applied. This cut eliminates
$\ksppn$ decays with successive $\pi^0$ decays in the Dalitz mode
($\pi^0 \rightarrow ee\gamma$) or with a $\gamma \rightarrow ee$
conversion in the material,
but reduces the signal acceptance by about a factor of two.
The main task of these searches, to control the background to a
sufficient precision, has been performed with minimal help of Monte Carlo
simulations. The main background in both modes, accidental
coincidences due to high intensity of the beam, could be estimated to
a level much lower than one event due to the availability
of a large ($>200$ ns) readout window. This allows one 
to have more than an order of
magnitude larger control region with respect to the coincidence signal
window of 3 ns. The contribution of $\kleegg$ background to the
$\kspee$ channel was estimated using 2001 data containing ten times
larger $K_L$ flux than the near-target data of 2002.

From the observed events, following branching fractions have been
extracted:
\begin{align}
BR(\kspee) &= (5.8^{+2.8}_{-2.3 stat} \pm 0.8_{syst})\times 10^{-9} \\
BR(\kspmm) &= (2.8^{+1.5}_{-1.2 stat} \pm 0.2_{syst})\times 10^{-9}
\end{align}
In the case of $\kspee$ the extrapolation to the full $m_{ee}$
kinematic region has been done assuming no form factor and the
uncertainty connected to this has been taken into account. In
principle, having measured both decays, the form factor, parametrised
as $W_S \sim a_S + b_S (m_{ll}/m_K)^2$, should be fully determined, at
least up to a sign ambiguity. Due to low statistics, however, only $|a_S|$
can be calculated using the the vector-meson-dominance (VMD) model
ansatz $b_S/a_S = 0.4$. Averaging both decay modes one obtains
 $|a_S|_{ll} = 1.21^{+0.22}_{-0.18}$.
The ratio
\begin{equation}
\frac{BR(\kspmm)}{BR(\kspee)} = 0.50 \pm 0.33
\end{equation}
is consistent with the prediction of the VMD model: 0.23~\cite{deip}. 

With these measurements, the last missing component 
in the effort to predict the branching fraction of decays $\klpll$ has
been determined. Following the theoretical analyses~\cite{bdi,isu}
the expected $BR$'s using SM fits are:
\begin{align}
BR(\klpee)_{SM} &\approx (17_{ICPV} \pm 9_{INTF} + 4_{DCPV})\times 10^{-12} \\
BR(\klpmm)_{SM} &\approx (8_{ICPV} \pm 3_{INTF} + 2_{DCPV} + 5_{CPC})\times 10^{-12}
\end{align}
where the uncertainties on the individual contributions
are at the level of 20-30\%. The sign of the interference term (INTF)
depends on the unknown sign of the $a_S$ parameter. 
A constructive interference is favoured now by
two theoretical groups \cite{bdi,fgd}. 
This means that despite of dominance of the
long-distance indirect CP-violating contribution relatively good
sensitivity to the $Im \lambda_t$ is retained through the
interference term.

The best present limits on the $BR$'s of $\klpll$ decays are provided
by the KTeV collaboration:
\begin{align}
BR(\klpee) &< 2.8 \times 10^{-10} (90\% CL) \cite{klpee} \\
BR(\klpmm)) &< 3.8 \times 10^{-10} (90\% CL) \cite{klpmm}
\end{align}
These limits are still far from SM predictions, however, many NP scenarios
foresee strong enhancements in the decay rates of these 
processes~\cite{np}. In
particular, a rather recent model of enhanced electroweak penguins
predicts an order of magnitude enhancement of the direct CP-violating
component in both decays~\cite{npa}. 
Therefore, even slight improvement of these
limits can have selective power on models beyond SM. Although
currently no
experiment plans to search for $\klpll$ decays, possibilities to
reach SM sensitivity have been studied and proposed~\cite{aug}.

\subsection{$\kcpnn$}

The decay $\kcpnn$ is fully dominated by short-distance
processes. The decay amplitude depends on both $Re
\lambda_t$ and $Im \lambda_t$. This, combined with one of other FCNC
kaon decays, offers a possibility of fully determining the CKM
unitarity triangle using exclusively rare kaon decays. 
A non-negligible charm admixture in the loop
contributes to slightly higher theoretical uncertainty compared to the
decay $\klpnn$. The SM prediction for the branching fraction is
\begin{equation}
BR(\kcpnn)_{SM} = (8.0 \pm 1.1) \times 10^{-11} \cite{npa,isid}
\end{equation}

The experiment E787 using the AGS at BNL 
searched for the decay $\kcpnn$ between years 1995 and 1999.
It used a separated $K^+$ beam stopped in an active
target. The signal signature included identification of the $K^+$ in a 
Cherenkov counter, measurement of the energy, momentum and range of
the charged track and the absence of any other signals in coincidence with
the examined event. In the kinematic region $211<P_{\pi}<229$ MeV/$c$
(Region I), where
$P_{\pi}$ is the $\pi^+$ momentum,
two events have been found with $0.14 \pm 0.05$ events
background expectation. The background determination was based
entirely on data~\cite{e787}.

The successor of the E787 experiment, E949, started to operate in the
year 2002. This experiment uses the same apparatus and technique as
E787 with upgraded photon veto system and range
stack. The upgrades allowed the experiment to run at higher
instantaneous intensity, with smaller dead time and should improve the
sensitivity in the less clean kinematic region $P_{\pi}<195$ MeV/$c$ (Region
II)~\cite{e787b}. 
First results from the 2002 run has been announced recently. One
more event was found with the background expectation of $0.30
\pm 0.03$ events~\cite{e949}. This new event has significantly lower signal to
background ratio compared to two E787 events. The combined
experimental branching ratio is
\begin{equation}
BR(\kcpnn) = (14.7^{+13.0}_{-8.9}) \times 10^{-11}
\end{equation}
This result is compatible with SM prediction. However, the fact that
the central value is about factor two above the expectation and the
uncertainties are large, leaves an exciting window for NP models~\cite{isid}.
The experiment E949 may take more data in the
future and improve the present uncertainty by a factor two or three.

\subsection{$\klmm$}

The decay $\klmm$ is the best measured among the rare kaon FCNC
processes with $BR(\klmm)=(7.18 \pm 0.17)\times 10^{-9}$~\cite{e781}. 
The short-distance process depends on $Re \lambda_t$ (or
$\rho$ in the Wolfenstein parametrisation) and, within SM, 
is estimated to contribute
with a $BR(\klmm)^{SD}_{SM}=(0.8\pm 0.3)\times 10^{-9}$~\cite{npa}.
Here the difficulty resides in determination of the
dominant long-distance contribution which proceeds through two photon
exchange. 

The absorptive part of the long-distance contribution can be
determined with the help of the $\klgg$ decay. Recently, two new
results were published:
\begin{align}
\frac{\Gamma(\klgg)}{\Gamma(\klppp)} 
     &= (2.81 \pm 0.01_{stat} \pm 0.02_{syst}) \times 10^{-3} 
  \mathrm{\ by\ NA48\ \cite{ksgg}} \\
\frac{\Gamma(\klgg)}{\Gamma(\klppp)} 
     &= (2.79 \pm 0.02_{stat} \pm 0.02_{syst}) \times 10^{-3} 
  \mathrm{\ by\ KLOE\ \cite{kloegg}}
\end{align}
Both results agree well with each other. 

More problematic is the small dispersive part of the long-distance
amplitude which interferes with the short-distance process. This part
can be related to the form factor in two photon decays, 
$\klgsg$ and $\klgsgs$, with one or both photons off shell. In an
earlier theoretical analysis this form factor was parametrised by a single
parameter $\alpha_{K^*}$~\cite{bms}. Later, refined analysis,
 introduced
a pair of parameters $\alpha, \beta$~\cite{dip}. 
The latest results from the KTeV collaboration, with notably a very
precise measurement of the $\kleeg$ decay~\cite{kleeg}, 
provide a new precision step in the
determination of this contribution. In the formalism of~\cite{bms}
the result is
$\alpha_{K^*}= -0.186 \pm 0.014$ which can be translated to $\alpha =
-1.611 \pm 0.044$ in terms of~\cite{dip}. The statistics available in
$\klgsgs$ modes, like $\kleemm$ and $\kleeee$ is not yet sufficient to
extract the parameter $\beta$. The $\kleeg$ result by KTeV agrees well
with their earlier result on $\klmmg$~\cite{klmmg} 
while it is in a 2.6 $\sigma$
discrepancy with the published NA48 result based on a small part of
the total data sample~\cite{na48eeg}. 
New result by NA48 with significantly higher statistics
is expected very soon\footnote{Shortly after this conference NA48
  presented a preliminary result using full 
statistics~\cite{leli} which is in
  agreement with KTeV results.}.

\subsection{Summary and Prospects}

\begin{figure}[htb]
\includegraphics[width=13cm]{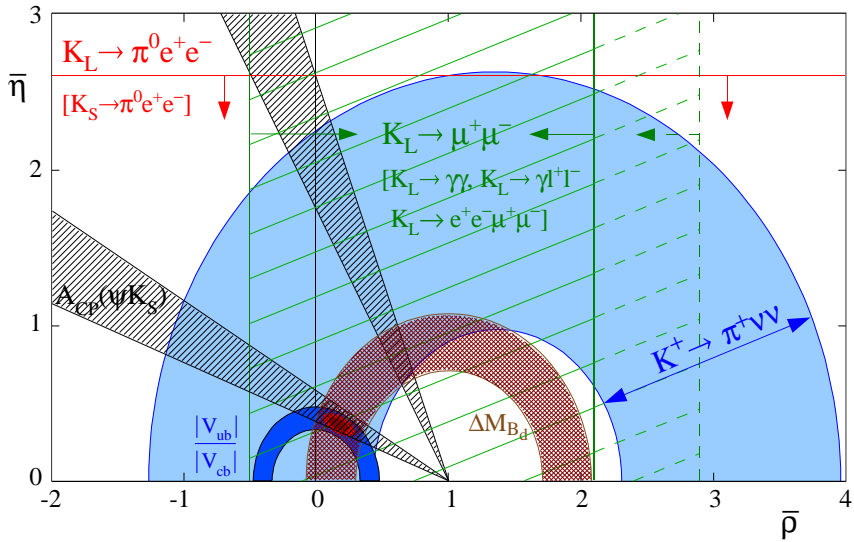}
 \caption{\it
      Comparison of present constraints on $\overline{\rho}$ and 
$\overline{\eta}$ from rare kaon decays and B-decays~\cite{isiunt}.
    \label{rhoeta} }
\end{figure}

A comparison of constraints imposed by the latest rare kaon decay results
and constraints provided by B-physics is shown in the
Fig.~\ref{rhoeta}~\cite{isiunt}. 
The impact of the latest E949 $\kcpnn$ event is not
shown but the change, due to small weight of this event is
small. The smaller $\overline{\rho}$ interval from $\klmm$
corresponds to the case of negative interference between short- and
long-distance contributions. The kaon constraints
are still about an order of magnitude away from SM expectations. On the
other hand, as mentioned before, many NP models predict substantial
enhancements of FCNC amplitudes. Therefore any improvement of these
limits can already have selective power among NP models. 

Even in the absence of a large NP signal, currently pursued experimental
projects provide a very good potential to reach the SM level in the
coming decade. The first experiment
dedicated to the search for
$\klpnn$ decay, 
the experiment E391a at KEK PS in Japan,
took first data this year and might double the statistics next
year. The technique is based on a low-energy pencil beam with highly
efficient photon veto system.
The aim is to reach
a single event sensitivity of about $4\times 10^{-10}$~\cite{e391a}
which is very close to some NP predictions~\cite{npa}. This experiment
is a pilot experiment for a much more ambitious project at the future
J-PARC facility which plans to collect more than 100 ``SM'' events.

Another experiment dedicated to the $\klpnn$ search, KOPIO, 
is under development at BNL with an approved budget for the year
2005. The experiment plans to use micro-bunched proton beam from AGS
extracting low-energy $K_L$ at a large angle from the target. The kaon
energy is measured by time of flight and the event reconstruction
relies on high redundancy of measured quantities~\cite{kopio}. The
expected sensitivity is similar to the Japanese 
project at J-PARC.
 
New projects to measure the $\kcpnn$ decay are also under way. In
addition to the experiment E949 at BNL, high statistics experiments are being
developed and proposed at Fermilab and CERN. The Fermilab CKM project
aims to collect about 100 ``SM'' events~\cite{ckmexp}. 
Unfortunately, at present the experiment is not supported by Fermilab. 
At CERN, a working group within the NA48 collaboration has studied the 
possibility to
use part of the NA48 equipment for a future $\kcpnn$ experiment under
the working name NA48/3. An
Expression of Interest has been submitted a few months
ago~\cite{na483} and a Letter of Intent will be submitted soon.

\section{Semileptonic Decays}
\label{semi}

The 2002 edition of PDG has revealed a slight deviation from the
unitarity in the first row of the CKM matrix~\cite{pdga}. The expression 
$|V_{us}|^2 + |V_{ud}|^2 + |V_{ub}|^2$, where the $V_{ub}$ is
negligible, deviated from unity by 2.2 standard deviations. Both
$V_{us}$ and $V_{ud}$ contribute with similar uncertainties.

This apparent discrepancy triggered an avalanche of experimental and
theoretical results, especially concerning $V_{us}$. The most
precise way of extracting the coupling
$V_{us}$ is by measuring the decay rate of 
the semileptonic kaon decay decay $\kpln$ ($K_{l3}$) 
and using the relation:
\begin{equation}
\Gamma(K_{l3}) \sim |V_{us}|^2 f_{+}^2(0) I^l_K (1+\delta^l_{rad})
\end{equation}
where $f_+(0)$ is the form factor normalisation, $I^l_K$ is the phase
space integral and $\delta^l_{rad}$ represents the radiative
corrections. More explicit formulae can be found elsewhere 
\cite{pdga,cpt6a,andre}. 
$\Gamma(K_{l3})$ and $I^l_K$
are measured experimentally, while $f_+(0)$ and 
$\delta^l_{rad}$ are calculated by
theorists. New development in all four components have been made recently.

The experiment KTeV at Fermilab, which was designed to measure
precisely the direct CP-violation parameter $Re\
\varepsilon'/\varepsilon$ and to search for rare $K_L$ decays, has
published recently a set of measurements of all the main $K_L$ decay
modes~\cite{ktevkla}. 
Measurements of five ratios of decay rates are summarised in
the Tab.\ref{ktevkl}. The ratios are chosen such that a large part of
the systematic effects, like e.g. trigger or detector efficiencies,
cancel. Also, in $\klpmn$ the
information from the muon system is not used and  in 
$\klpppc$ the $\pi^0$ is not reconstructed to minimise systematic biases.
Main systematic uncertainties originate from detector imperfections 
difficult to simulate, in particular the reconstruction
efficiencies and material simulation. Clearly, the most difficult
ratio $\Gamma(\klppp)/\Gamma(\klpen)$, where information from
different detectors has to be combined, suffers from the largest uncertainty.
Using the five ratios all main six $K_L$ branching ratios can be extracted.

\begin{table}
\centering
\caption{ \it KTeV measurements of main $K_L$ decay widths. The first
  uncertainty is statistical and the second systematic.
}
\vskip 0.1 in
\begin{tabular}{l|r} \hline
Decay Modes & Partial Width Ratio \\ \hline
$\Gamma(\klpmn)/\Gamma(\klpen)$  & $0.6640\pm 0.0014 \pm 0.0022$ \\
$\Gamma(\klppp)/\Gamma(\klpen)$  & $0.4782\pm 0.0014 \pm 0.0053$ \\
$\Gamma(\klpppc)/\Gamma(\klpen)$ & $0.3078\pm 0.0005 \pm 0.0017$ \\
$\Gamma(\klpp)/\Gamma(\klpen)$   & 
         $(4.856\pm 0.017 \pm 0.023)\times 10^{-3}$ \\
$\Gamma(\klppn)/\Gamma(\klppp)$  & 
         $(4.446\pm 0.016 \pm 0.019)\times 10^{-3}$ \\
\hline
\end{tabular}
\label{ktevkl}
\end{table}

All ratios involving the $\klpen$ ($K_{e3}$) decay disagree by several
standard deviations from the present world averages~\cite{pdgb}. 
Therefore, new
measurements of these decay widths are highly desirable and both 
KLOE and NA48 are in a process of analysing these decays.

Together with the $K_{l3}$ decay widths, KTeV measured also the form factor
shapes extracting for the first time the quadratic term
$\lambda_+''$~\cite{ktevklb}. 
Combining both $\klpen$ and $\klpmn$ results and using
an in-house calculation of radiative corrections~\cite{andre} they
obtain~\cite{ktevklc}:
\begin{equation}
|V_{us}| f_+(0) = 0.2165 \pm 0.0012
\end{equation}

Another interesting result addressing the $V_{us}$ determination has
been recently presented by the KLOE collaboration. This unique
experiment collects $\phi \rightarrow K_S K_L$ data at the Frascati
$\phi$-factory DAPHNE. By tagging the opposite 
$K_L$ this experiment can access $K_S$ decay modes which, in fixed
target experiments, would be flooded by $K_L$ background. The
$\kspp{(\gamma)}$ background is rejected by requiring $m_{\pi\pi}<490$
MeV/c$^2$. The remaining background is separated by 
testing for the presence of neutrino comparing the
missing energy and missing momentum, where the original direction of
$K_S$ is determined from the position of the opposite $K_L$. The
branching fraction is 
$BR(\kspen)=(7.09 \pm 0.07_{stat} 
\pm 0.08_{syst})\times 10^{-4}$~\cite{kloevus}. 
Using the
CKM working group recipe~\cite{batt} they obtain
\begin{equation}
|V_{us}| f_+(0) = 0.2157 \pm 0.0018
\end{equation}

New measurement of the $\klpen$ decay rate and form factor 
has been presented by NA48
at~\cite{kk}. This measurement is normalised to an inclusive 2-track
decay rate:
\begin{equation}
R=\frac{\Gamma(\klpen)}{\Gamma(K_L\rightarrow 2 track)} = 0.497 \pm 0.004
\end{equation}
To extract the branching fraction $BR(K_L\rightarrow 2 track)$
is calculated by subtracting
from unity all major neutral decay modes. Here the dominant
contribution comes from $\klppp$ decay. At present, the new KTeV
result on $BR(\klppp)$~\cite{ktevkla} 
disagrees by six standard deviations from the PDG value~\cite{pdgb}. 
A new measurement of $BR(\klppp)$ is needed to resolve this 
situation\footnote{Shortly after this conference a new preliminary
  result has been presented by NA48~\cite{leli, rainer} which confirms the KTeV
  measurement of $BR(\klppp)$.}. 

\begin{figure}[htb]
\includegraphics[width=13cm]{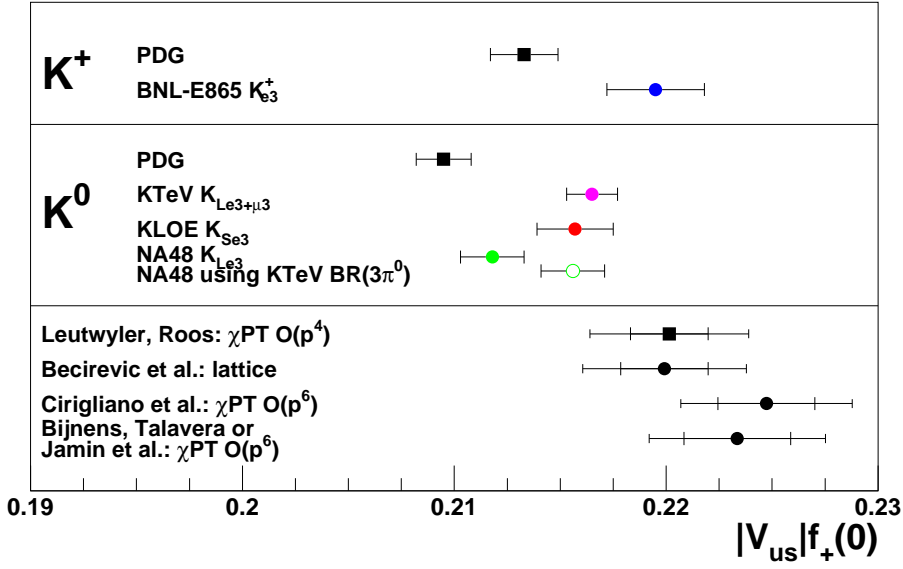}
 \caption{\it
    Comparison of old (squares) and new (circles) results on
    $V_{us}$. $K^+$ results are corrected by the ratio of $f_+(0)$ for
    neutral and charged kaons.
    The theoretical results correspond to
    $f_+(0)(1-|V_{ud}|^2)^{1/2}$ where the larger error bar includes
    the uncertainty on $V_{ud}$. 
    \label{vus} }
\end{figure}

In the last months, several new theoretical calculations of $f_+(0)$
have been published. A new lattice calculation~\cite{latt} agrees with
the original $\cpt$ calculation at $O(p^4)$~\cite{cpt4} while
$\cpt$ calculations at $O(p^6)$ obtain slightly higher 
value~\cite{cpt6a,cpt6b}.
A summary of the present experimental and
theoretical situation is shown in the Fig.~\ref{vus}. In general the
experimental results from semileptonic decays of neutral kaons 
are consistent with each other and 
with the earlier measurement of $\kcpen$ by 
E865~\cite{e865}. However, the situation, 
especially from the theoretical side, is not completely
satisfactory. 

Two other new results falling into the category of semileptonic
decays have been published by NA48. The decay $\klppen$, interesting
e.g. for the study of the $\pi\pi$ scattering,  has been
measured with an improved precision \cite{laur}:
\begin{equation}
BR(\klppen) = (5.21 \pm 0.07_{stat} \pm 0.09_{syst})\times 10^{-5}
\end{equation}
A set of form factors was fitted using distributions of five
Cabibbo-Maksymowicz variables. The result agrees with previous
measurements and with theoretical predictions \cite{abt}.

The second result is the measurement of the radiative $\klpeng$
decay. Here, the analysis is complicated by the fact that the acceptance
calculation can depend on the theoretical model inserted 
into the Monte Carlo
event generator. NA48 has performed a model-independent analysis by
re-weighting the Monte Carlo events according to kinematic
distributions obtained from data. The result \cite{klpeng}
\begin{equation}
\frac{\Gamma(\klpeng)}{\Gamma(\klpen)} = 
   (9.60 \pm 0.07_{stat}\ ^{+0.12} _{-0.11syst})\times 10^{-3}
\end{equation}
agrees with theoretical predictions based on $\cpt$ \cite{klpengt} but
disagrees with a published result from KTeV which has similar statistical
precision \cite{klpenge}.

\section{CP Violation}

The decay $\ksppp$ is purely CP-violating and is, up to now,
unobserved. The CP-violation parameter $\eta_{000}\equiv
A(\ksppp)/A(\klppp)$ is proportional to the expression $\varepsilon +
i ImA_1/ReA_1$, where $\varepsilon$ is the parameter of indirect CP
violation in the kaon system 
and $A_1$ is the isospin 1 decay amplitude. The imaginary
part of $\eta_{000}$ is in principle sensitive to direct CP
violation \cite{pdgb}. 
In addition, the poor knowledge of $\eta_{000}$, limits at
present the CPT tests in the K system.

In a recently published paper, NA48 searches for the decay $\ksppp$ by
exploring the interference between $K_S$ and $K_L$ and fitting the
decay distribution of the data taken during the special near-target
run in the year 2000. The result~\cite{ksppp}
\begin{align}
Re \eta_{000} &= -0.002 \pm 0.011_{stat} \pm 0.015_{syst} \\
Im \eta_{000} &= -0.003 \pm 0.013_{stat} \pm 0.017_{syst}
\end{align}
represents an order of magnitude improvement with respect to previous
results and leads to a limit $|\eta_{000}|<0.045$ at 90\% CL level.
This can be
translated to $BR(\ksppp) < 7.4 \times 10^{-7}$
which is more than 
an order of
magnitude better than the currently published limit but still two orders of
magnitude above the SM expectation. Using the Bell-Steinberger~\cite{bs}
relation and other measured kaon CP-violating decay amplitudes 
one can extract a significantly improved limit on 
the CPT-violating parameter $\delta$: $Im \delta = (-0.2 \pm 2.0) \times
10^{-5}$. Assuming that CPT is conserved in the decay leads to a
measurement of $K^0\overline{K^0}$ mass difference 
$m_{K^0}-m_{\overline{K^0}} = (-0.2\pm 2.8) \times 10^{-19}$ GeV/$c^2$.
By imposing full CPT invariance the limit on branching fraction
becomes $BR(\ksppp) < 2.3 \times 10^{-7}$.

Similar results have been recently presented by the KLOE
collaboration however, obtained by completely different
technique. KLOE experiment is capable to search for the decay
$\ksppp$ directly by tagging the $K_S$ decays with an opposite $K_L$
as described in the section \ref{semi}. They observe 4 events in the
signal region with $3.2 \pm 1.4$ background events expected. This
leads to a limit $BR(\ksppp) < 2.1 \times 10^{-7}$  at 90\% CL or
$|\eta_{000}|<0.024$ \cite{kloeppp}. 
New data from KLOE promise steady improvements
of these limits in the future. 

Using their newly measured branching fractions of all main $K_L$
decays, KTeV collaboration obtained also a more precise measurement of
the CP-violating parameter $|\eta_{+-}|$.
With the help of the expression
\begin{equation}
|\eta_{+-}|^2=\frac{\Gamma(\klpp)}{\Gamma(\kspp)} =
\frac{\tau_S}{\tau_L}\frac{BR(\klpp) + 
              BR(\klppn) (1+6 Re\ \varepsilon'/\varepsilon)}
                          {1 - BR(\kspen)}
\end{equation}
where the 
normalisation is done under the assumption $\Gamma(\kspen)=\Gamma(\klpen)$,
KTeV  obtains \cite{ktevkla}:
\begin{equation}
|\eta_{+-}| = (2.228 \pm 0.005_{KTeV} \pm 0.009_{\tau_{KL}}) \times 10^{-3}
\end{equation}
The main uncertainty comes from the poor knowledge of the 
$K_L$ lifetime, which may be
measured more precisely soon by KLOE \cite{franz}. 
This result differs by more than three
standard deviations from the current world average
\cite{pdgb}. 
This is
another indication of the inconsistencies in our understanding of
the main $K_L$ branching ratios, which will hopefully be resolved soon by new
results from KLOE and NA48.

Finally, new searches for direct CP-violation are under way in the
decays of charged kaons. The second extension of the NA48 experiment,
NA48/2 \cite{na482pr}, uses simultaneous $K^+$ and $K^-$ beams to measure the
asymmetry in the Dalitz plot slope parameter $g$ 
\begin{equation}
A_g \equiv \frac{g(\kpppp)-g(\kmppp)}{g(\kpppp)+g(\kmppp)}
\end{equation}
More than 2 billion $\kpmpppc$ and few hundred millions of $\kpmpppn$ have been
collected in years 2003 and 2004 \cite{kekel}. 
This will allow to access $A_g$ to
the level better than $2\times 10^{-4}$ which is better than an order
of magnitude with respect to previous experiments. 
The theoretical predictions within SM vary between $10^{-6}$ and $5 \times
10^{-5}$ \cite{agth} 
while enhancements to the level of $10^{-4}$ are predicted by
models beyond SM \cite{agth2}.
Similar sensitivity is expected from the experiment OKA in Protvino.

\section{Conclusions}

Most recent developments in the experimental kaon physics have been
reviewed. The amount of presented results, despite of the fact
that this review is necessarily selective and incomplete, documents
the unfading activity in this field. New results in rare decays have shed
more light on the FCNC processes which seem to be one of the best
windows to the New Physics in the future. Especially the golden modes
$\kpnn$ are the targets of new projects in kaon physics. They promise
an exciting interplay together with B-factories and Tevatron or LHC
experiments, in the coming years in the quest for physics beyond Standard
Model. 

The situation around the unitarity test in the first row of the CKM
mixing matrix seems to improve. The new experimental
results are consistent with each other. On the other hand various
theoretical calculations lead to different conclusions. The question
``Is the unitarity in CKM matrix conserved?'' is not completely
answered yet.

In the intermediate term, 
many new results are expected from experiments at Frascati,
Brookhaven, CERN and KEK. The luminosity of the DAPHNE machine is
improving and promises new interesting results from the powerful
experiment KLOE. The huge sample of charged kaon decays accumulated by
NA48/2 will yield not only new results on direct CP violation but also
on a full palette of rare decays of charged kaons. Of course, a continuing
support from the theorists is essential to keep kaon
physics attractive.


\begin{thebibliography}{99}
\bibitem{disc} J.H. Christenson {\it et al.}, \Journal{\PRL}{13}{1964}{138}.
\bibitem{ckm} M. Kobayashi and K. Maskawa, 
             \Journal{Prog. Theor. Phys}{49}{1973}{652}.
\bibitem{gim}  S.L. Glashow, J. Iliopoulos and L. Maiani, 
           \Journal{\PRD}{2}{1970}{1285}.
\bibitem{na31} G. Barr {\it et al.} (NA31), \Journal{\PLB}{317}{1993}{233}.
\bibitem{na48a} A. Lai {\it et al.} (NA48), \Journal{\EPJ}{22}{2001}{231}.
\bibitem{na48b} J.R. Batley {\it et al.} (NA48), 
           \Journal{\PLB}{544}{2002}{97}.
\bibitem{ktev} A. Alavi-Harati {\it et al.} (KTeV), \Journal{\PRD}
           {67}{2003}{012005}.
\bibitem{np} G. Colangelo and G. Isidori,
                                 \Journal{JHEP}{09}{1998}{009}, \\
             Y. Nir and M.P. Worah, \Journal{\PLB}{423}{1998}{319}, \\
             A.J. Buras, A. Romanino and L. Silvestrini, 
                        \Journal{\NPB}{520}{1998}{3}, \\
             A.J. Buras {\it et al.}, \Journal{\NPB}{566}{2000}{3}, \\
             G. D'Ambrosio and G. Isidori, \Journal{\PLB}{530}{2002}{108}, \\
             A.J. Buras {\it et al.}, \Journal{\NPB}{660}{2003}{225}, \\
             Y. Grossman, G. Isidori and H. Murayama, 
                          \Journal{\PLB}{588}{2004}{74}.
\bibitem{npa} A.J. Buras {\it et al.}, hep-ph/0402112.
\bibitem{litt} L.S. Littenberg \Journal{\PRD}{39}{1989}{3322}.
\bibitem{wolf} L. Wolfenstein \Journal{\PRL}{51}{1983}{1945}.
\bibitem{klpnn} A. Alavi-Harati {\it et al.} (KTeV), \Journal{\PRD}
           {61}{2000}{072006}.
\bibitem{isid} G. Isidori, hep-ph/0307014.
\bibitem{bdi} G. Buchalla, G. D'Ambrosio and G. Isidori, 
               \Journal{\NPB}{672}{2003}{387}.
\bibitem{isu} G. Isidori, C. Smith and R. Unterdorfer, hep-ph/0404127.
\bibitem{batt} M. Battaglia {\it et al.} ed., Workshop on CKM Unitarity
  Triangle, hep-ph/0304132.
\bibitem{klpgg} A. Lai {\it et al.} (NA48), \Journal{\PLB}{536}{2002}{229},\\
  A. Alavi-Harati {\it et al.} (KTeV), \Journal{\PRL}{83}{1999}{917}.
\bibitem{theoks} L.M. Sehgal, \Journal{\NPB}{19}{1970}{445}, \\
G. Ecker, A. Pich and E. de Rafael, \Journal{\NPB}{291}{1987}{692}, \\
 C. Bruno and J. Prades, \Journal{\ZPC}{57}{1993}{585}.
\bibitem{add2} J.R. Batley  {\it et al.} (NA48), CERN/SPSC 2000-002.
\bibitem{ksgg} J.R. Batley  {\it et al.} (NA48), 
                 \Journal{\PLB}{551}{2003}{7}.
\bibitem{kspgg} J.R. Batley  {\it et al.} (NA48), 
                 \Journal{\PLB}{578}{2004}{276}.
\bibitem{ksppp} A. Lai  {\it et al.} (NA48), 
                hep-ex/0408053 submitted to \PLB.
\bibitem{kspee} J.R. Batley  {\it et al.} (NA48), 
                 \Journal{\PLB}{576}{2003}{43}.
\bibitem{kspmm} J.R. Batley  {\it et al.} (NA48), 
                 \Journal{\PLB}{599}{2004}{197}.
\bibitem{deip} G. D'Ambrosio {\it et al.},
  \Journal{JHEP}{08}{1998}{004}.
\bibitem{fgd} S. Friot, D. Greynat and E. de Rafael,
          \Journal{\PLB}{595}{2004}{301}.
\bibitem{klpee} A. Alavi-Harati {\it et al.} (KTeV), 
          hep-ex/0309072.
\bibitem{klpmm} A. Alavi-Harati {\it et al.} (KTeV), \Journal{\PRL}
           {84}{2000}{5279}.
\bibitem{aug} A. Belyaev {\it et al.}, hep-ph/0107046, \\
              A. Ceccucci, \Journal{Int. J. Mod. Phys.A}{19}{2004}{889}.
\bibitem{e787} S. Adler  {\it et al.} (E787), \Journal{\PRL}{88}{2002}{041803}.
\bibitem{e787b} S. Adler  {\it et al.} (E787), 
                     \Journal{\PRD}{70}{2004}{037102}.
\bibitem{e949} V.V. Anisimovski  {\it et al.} (E949), hep-ex/0403036 .
\bibitem{e781} D. Ambrose  {\it et al.} (E781), 
                     \Journal{\PRL}{84}{2000}{1389}.
\bibitem{kloegg} M. Adinolfi  {\it et al.} (KLOE), 
                     \Journal{\PLB}{566}{2003}{61}.
\bibitem{bms} L. Bergstr\"om, E. Mass\'o and P. Singer, 
                     \Journal{\PLB}{131}{1983}{229}.
\bibitem{dip} G. D'Ambrosio, G. Isidori and J. Portol\'es,
                     \Journal{\PLB}{423}{1998}{385}.
\bibitem{kleeg} J. LaDue, ``Understanding Dalitz decays of the $K_L$,
  in particular the decays of $\kleeg$ and $\kleeee$'', PhD thesis,
  University of Colorado - Boulder, May 2003.
\bibitem{klmmg} A. Alavi-Harati {\it et al.} (KTeV), 
           \Journal{\PRL} {87}{2001}{071801}.
\bibitem{na48eeg} V. Fanti  {\it et al.} (NA48), 
                 \Journal{\PLB}{458}{1999}{553}.
\bibitem{leli} L. Litov (for the collaboration NA48),
  ``Measurement of $V_{us}$. Recent NA48 Results on Semileptonic and
  Rare Kaon Decays.'', talk given at ICHEP 2004, Beijing (August
  2004), 
{\tt http://ichep04.ihep.ac.cn/8\_cp.htm}.
\bibitem{isiunt} G. Isidori and R. Unterdorfer, hep-ph/0311084.
\bibitem{e391a} T. Inagaki, Seminar at Riken (July 2004) 
{\tt http://www-ps.kek.jp/e391/
pub/documents/ohp/riken-inagaki-2004.ppt},
\bibitem{kopio} I.-H. Chiang {\it et al.} (KOPIO), ``KOPIO - A study
  of $\klpnn$'', RSVP MRE proposal (November 1999).
\bibitem{ckmexp} P.S. Cooper {\it et al.} (CKM),
          \Journal{Nucl. Phys. Proc. Suppl. B}{99}{2001}{121}.
\bibitem{na483} NA48-Future Working Group, `` Expression of Interest
  to Measure Rare Kaon Decays at the CERN SPS'', CERN-SPSC-2004-010
  (April 2004).
\bibitem{cpt6a} V. Cirigliano, H. Neufeld and H. Pichl, hep-ph/0401173.
\bibitem{andre} T.C. Andre, hep-ph/0406006.
\bibitem{pdga} K. Hagiwara {\it et al.} (PDG), 
              \Journal{\PRD}{66}{2002}{010001}.
\bibitem{ktevkla} T. Alexopoulos {\it et al.} (KTeV),
           hep-ex/0406002.
\bibitem{pdgb} S. Eidelman {\it et al.} (PDG), \Journal{\PLB}{592}{2004}{1}.
\bibitem{ktevklb} T. Alexopoulos {\it et al.} (KTeV),
           hep-ex/0406003.
\bibitem{ktevklc} T. Alexopoulos {\it et al.} (KTeV),
           hep-ex/0406001.
\bibitem{kloevus} see e.g. C. Gatti (KLOE), 
   ``K Semileptonic Decays at KLOE'', talk given at 
DA$\Phi$NE 2004: Physics at meson factories, Frascati (June 2004), 
{\tt http://www.lnf.infn.it/conference/2004/dafne04}.
\bibitem{kk} K. Kleinknecht (NA48), ``$V_{us}$ from Semileptonic $K^0$
  decays'', talk given at Heavy Quarks and Leptons, San Juan (June 2004),
{\tt http://charma.uprm.edu/hql04}.
\bibitem{rainer} R. Wanke (NA48), ``Solving the unitarity question:
  New NA48 results on semileptonic Kaon and Hyperon decays'', CERN
  seminar (August 2004), {\tt
  http://agenda.cern.ch/fullAgenda.php?ida=a043410\#2004-08-10}. 
\bibitem{latt} D. Be\'cirevi\'c {\it et al.}, hep-ph/0403217.
\bibitem{cpt4} H. Leutwyler and M. Roos, \Journal{\ZPC}{25}{1984}{91}.
\bibitem{cpt6b} J. Bijnens and P. Talavera,
                \Journal{\NPB}{669}{2003}{341}, \\
                M. Jamin, J.A. Oller and A. Pich, hep-ph/0401080.
\bibitem{e865} A. Sher {\it et al.} (E865), \Journal{\PRL}{91}{2003}{261802}.
\bibitem{laur} J.R. Batley  {\it et al.} (NA48), 
              \Journal{\PLB}{595}{2004}{75}.
\bibitem{abt} G. Amor\'os, J. Bijnens and P. Talavera,
            \Journal{\NPB}{585}{2000}{293}.
\bibitem{klpeng} see e.g. M. Slater (for the collaboration NA48),
  ``Latest Results from NA48 and NA48/1'', talk given at 39th
  Rencontres de Moriond on Electroweak Interactions and Unified
  Theories, La Thuile (March 2004),
  hep-ex/0406064.
\bibitem{klpengt} H. Fearing {\it et al.},
                        \Journal{\PRD}{2}{1970}{542}, \\
                  M.G. Doncel, \Journal{\PLB}{32}{1970}{623}.
\bibitem{klpenge} A. Alavi-Harati {\it et al.} (KTeV), 
           \Journal{\PRD} {64}{2001}{112004}.
\bibitem{bs} J.S. Bell and J. Steinberger, Proc. of Oxford
  Int. Conf. on Elementary Part. (1965) 195.
\bibitem{kloeppp} see e.g. M. Martini (KLOE), 
   ``A Direct Search of $K_{S}$ to 3$\pi^0$ Decay with the KLOE Detector'', 
talk given at 
DA$\Phi$NE 2004: Physics at meson factories, Frascati (June 2004), 
{\tt http://www.lnf.infn.it/conference/2004/dafne04}.
\bibitem{franz} P. Franzini (KLOE), ``Kaon decays and $V_{US}$'',
  talk given at this conference, hep-ex/0408150.
\bibitem{na482pr} R. Batley {\it et al.} (NA48), CERN/SPSC/2000-003,
             CERN/SPSC/2003-033.
\bibitem{kekel} V. Kekelidze, ``Charged Kaon Experiment at CERN SPS -
  NA48/2'', talk given at ICHEP 2004, Beijing (August 2004), {\tt
  http://ichep04.ihep.ac.cn/8\_cp.htm}. 
\bibitem{agth} L. Maiani and N. Paver, The second DA$\Phi$NE Physics
  Handbook, INFN, LNF, Vol 1. (1995) 51, \\
           E.P. Shabalin, hep-ph/0405229, \\
           A.A. Belkov, A.V. Lanyov and G. Bohm, hep-ph/0311209.
\bibitem{agth2} G. D'Ambrosio, G. Isidori and G. Martinelli, 
                \Journal{\PLB}{480}{2000}{164}.

\end{thebibliography}
\end{document}